\documentclass{article} 
\usepackage{nips14submit_e,times}
\usepackage[numbers,sort&compress]{natbib}
\usepackage{hyperref}
\usepackage{url}
\usepackage{amssymb}
 \usepackage{graphicx} 
 \usepackage{comment}
   \usepackage{amsmath,amsthm, amssymb, latexsym}

\usepackage{mydefs}

\title{Sneaky light stop}

\author{Till Eifert\and Benjamin Nachman}

\author{
Till Eifert\\
CERN\\
\texttt{eifert@cern.ch} \\
\And
Benjamin Nachman\\
SLAC, Stanford University\\
\texttt{bnachman@cern.ch} \\}

\newcommand{\ttbar}{\ensuremath{t\bar{t}}}
\newcommand{\tone}{\ensuremath{{\tilde{t}^{}_{1}}}}

\newcommand{\stoppair}{\ensuremath{\tilde{t}\tilde{t}^*}}

\def\topLSP{\ensuremath{\tone \to t \ninoone}}
\def\threeBody{\ensuremath{\tone \to b W \ninoone}}

\nipsfinalcopy 

\begin{document}

\maketitle

\begin{abstract}
A light supersymmetric top quark partner (stop) with a mass nearly degenerate with that of the Standard Model (SM) top quark can evade direct searches.
The precise measurement of SM top properties such as the cross-section has been suggested to give a handle for this `stealth stop' scenario. 
We present an estimate of the potential impact a light stop may have on top quark mass measurements. 
The results indicate that certain light stop models may induce a bias of up to a few \GeV,
and that this effect can hide the shift in, and hence sensitivity from, cross-section measurements.
The studies make some simplifying assumptions for the top quark measurement technique, and are based on
truth-level samples.
\end{abstract}

\section{Introduction}

Naturalness arguments suggest that if supersymmetry (SUSY) provides a solution to the hierarchy problem, then the supersymmetric partner of the top quark, the stop (\stop), should be relatively light.  An experimentally difficult, but well motivated region is when 
the mass of the lighter stop (\tone) is nearly mass-degenerate with that of the SM top quark, 
$m_{\tone} \sim m_t$ (this will be refereed to as a degenerate stop).\footnote{
We focus on the three-body (\threeBody) and two-body (\topLSP) decay processes, assuming that the lightest neutralino (\ninoone) is the lightest SUSY particle and that $R$-parity is conserved. The signal from a degenerate \tone\ that decays via other SUSY particles is typically well covered by direct searches. 
}
Recent studies (see for example refs.~\cite{atlasxs,theoryxs}) have shown that there is some sensitivity to a degenerate stop via precision measurements of top quark properties, such as the \ttbar\ cross-section ($\sigma_{\ttbar}$).  We make the simple observation that measurements which exploit the cross-section could be effected by a bias in the top measurement due the presence of a light stop.  In particular, since $\sigma_{\ttbar}$ increases with decreasing top quark mass, a negative shift in the measured top quark mass would increase the predicted $\ttbar$ cross-section and could hide the additional contribution to the measured cross-section from direct stop pair production.

Exploiting precision measurements of the \ttbar\ cross-section is due in part to the NNLO+NNLL precision~\cite{ttbar-xsection} that reduces the theoretical uncertainty in $\sigma_{\ttbar}$ to about 5\%, which is sensitive to the $\mathcal{O}(10\%)$ contribution of a degenerate stop.
The recent ATLAS cross-section measurement~\cite{atlasxs}, with a measurement uncertainty of about 4\%, includes an interpretation that sets 95\% CL exclusion limits on a stop in the range $m_t<m_{\tone} < 177$\,\GeV\ for a 100\% branching ratio of \topLSP, a nearly massless \ninoone, and for a \tone\ that is mostly the partner of the stop-right.  
Another interesting approach, as pointed out for instance in ref.~\cite{spin-correlation}, is to exploit the difference in spin of the stop and top quark. 
ATLAS has recently released preliminary results where a degenerate stop is searched for using both the \ttbar\ cross-section and spin-related kinematic informationf in the azimuthal angle between the two charged leptons~\cite{atlas-spin}.  A light stop decaying with a branching ratio of 100\% via \topLSP\ is excluded at 95\% CL in the range from the top quark mass up to 191\,\GeV, for the same \tone\ assumptions as in the ATLAS exclusion from the cross-section measurement.
The sensitivity degrades by $30$\% without the cross-section constraint.

We explore how the presence of a `sneaky' light stop could be hidden from these and future measurements due to a shift in the measured top quark mass.  Section~\ref{sec:method} first introduces a simple mass measurement technique which is used in  section~\ref{sec:results} to show how a degenerate stop can bias precision top quark measurements.  Section~\ref{sec:conclusions} summarizes the implications for current and future measurements.

\section{Method}\label{sec:method}

The top quark mass is measured  from the event kinematics using various experimental techniques. These techniques consider the various degrees of per-event kinematic constraints available for the
zero, one, and two-lepton (electron or muon) channels of the \ttbar\ decay, and typically perform in-situ calibration of some quantities (such as the effective jet energy scale). Another type of measurement of the top quark mass is performed using topologies enhanced in single-top events.
A recent overview can be found in ref.~\cite{TOP2014}. The most precise measurements are obtained using the one-lepton channel (also referred to as lepton+jets channel) 
and measure $m_{jjj}$, the invariant mass of the three jets associated with the hadronic top decay.\footnote{These are measurements of the {\it Monte Carlo} mass, which is related to a well-defined QFT top quark mass within ambiguities of $\mathcal{O}{(\Lambda_\text{QCD})}$ or more, see e.g. ref.~\cite{svenMoch_topAmbiguity}.  For our purposes, this is not an important detail as the corresponding uncertainty is included in the theoretical cross-section. 
}  
The results of the two single-measurements with the highest precision as of today are 
$m_t = 174.98 \pm 0.76$\,\GeV\ and $172.04 \pm 0.77$\,\GeV\
from the D0~\cite{D0_mass} and CMS~\cite{cmsmass} Collaborations, respectively.
We will focus on this type of measurement in the one-lepton channel, and 
use a rough approximation of the method (described in the following) to study the potential bias in the measured top quark mass in the presence of a light stop. 
The potential bias of other top quark mass measurements and techniques requires dedicated studies. We leave the investigation of this question to future work.\footnote{
While the measurement in the zero-lepton channel might have a rather similar bias as the one-lepton channel (both measure $m_{jjj}$), techniques in the two-lepton channel
that exploit kinematic edges might turn out to be robust, albeit they have have less sensitivity than the one-lepton channel.}

We simulate \ttbar\ and direct \tone\ pair production using {\sc Herwig++ 2.7}~\cite{herwig,herwig2}. For the latter, we consider both the two-body \topLSP\ decays for $m_{\tone}>m_t$ and three-body \threeBody\ decays for $m_{\tone}<m_t$.\footnote{
The separation into two- and three-body decays is not strict due to per-event variations with the natural widths of the top and stop.}
Finite width effects in the simulation of three-body decays are taken into account~\cite{offshell}. 
No detector simulation is performed. We consider proton--proton collisions at a centre-of-mass energy of $\sqrt{s} = 8$\,\TeV\ and $14$\,\TeV\ (LHC8 and LHC14, respectively) and proton--antiproton collisions at $\sqrt{s} = 1.96$\,\TeV\ (Tevatron).
The \ttbar\ events are normalized using theoretical cross-sections at NNLO+NNLL~\cite{ttbar-xsection} precision for both the two LHC and the Tevatron settings. 
The values for a reference top quark mass of $m_t = 172.5$\,\GeV\ are $253$\,pb (LHC8), $832$\,pb (LHC14), and $7.4$\,pb (Tevatron) 
as obtained using {\sc top++2.0}~\cite{top++2.0} and with
the PDF4LHC prescription~\cite{PDF4LHC} (LHC8 and LHC14) and the {\sc MSTW2008nnlo68cl}~\cite{MSTW2008nnlo68cl} PDF set (Tevatron). Variations in the \ttbar\ cross-section as a function of $m_t$ are obtained using the reference values above together with an accurate $m_t$ parametrization described in ref.~\cite{top-mass-parametrization}.
The SUSY stop samples are normalized using theoretical cross-sections at NLO+NLL precision for the LHC8~\cite{8TeVstopxs}, LHC14~\cite{14TeVstopxs}, and Tevatron~\cite{susyxsaTevatron} settings.\footnote{
The k-factor from NLO+NLL to NNLO+NNLL for the SM \ttbar\ process is at the per cent-level (see ref.~\cite{NNLO-kfactor}). Hence, applying this k-factor to the stop signal (in order to treat both processes on the same footing) would not change the results.
} The LHC8 and LHC14 \tone\ pair production cross-sections are provided with a fine granularity in $m_{\tone}$, while for the Tevatron the $m_{\tone}$ variations are obtained following the approach described in ref.~\cite{stopxs}.
For comparison, the values for a stop quark mass of $m_{\tone} = 175$\,\GeV\ are $36.8$\,pb (LHC8), $143.4$\,pb (LHC14), and $0.70$\,pb (Tevatron).

\begin{figure}[h!]
\begin{center}
\includegraphics[width=0.5\textwidth]{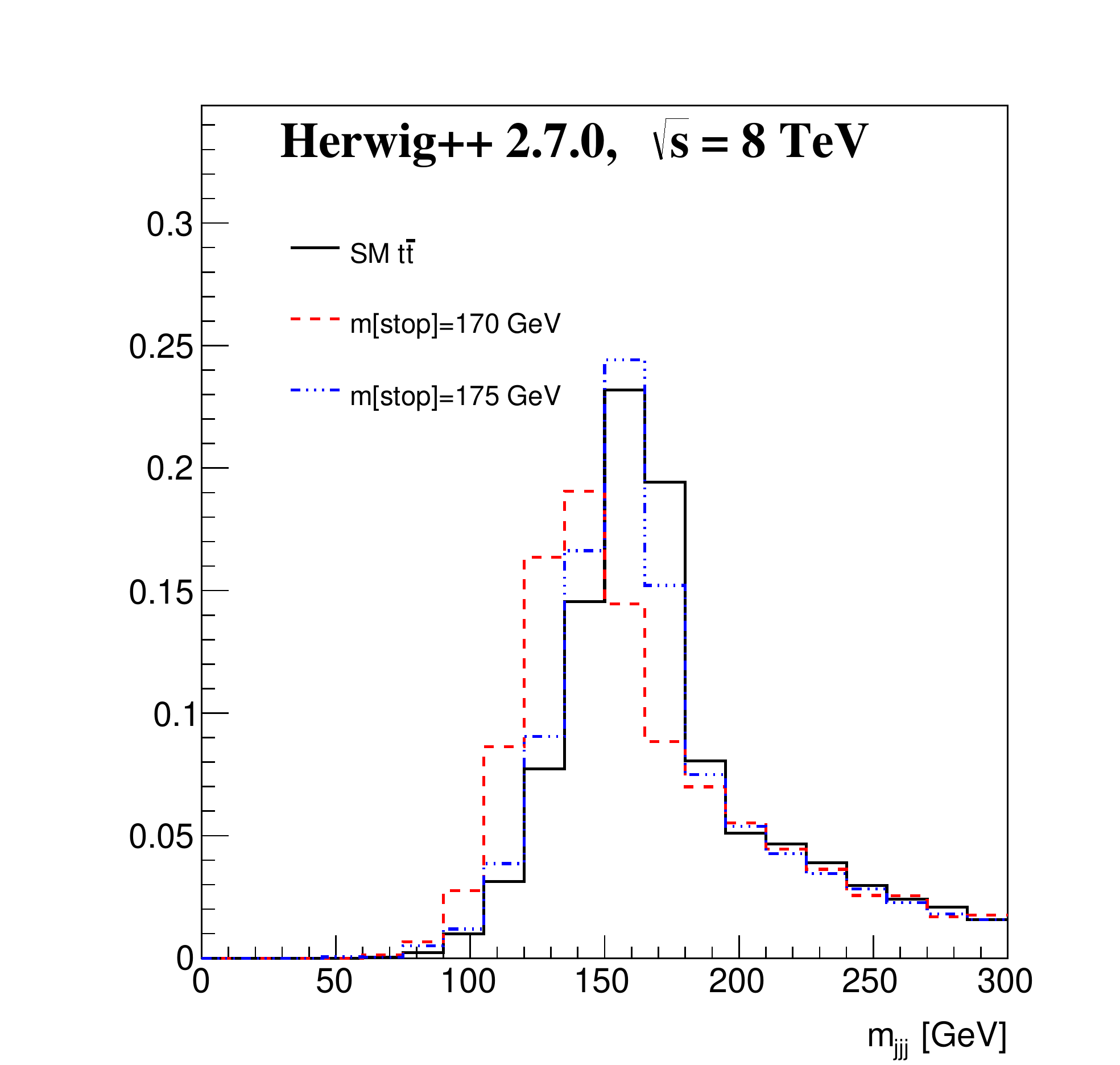}\includegraphics[width=0.5\textwidth]{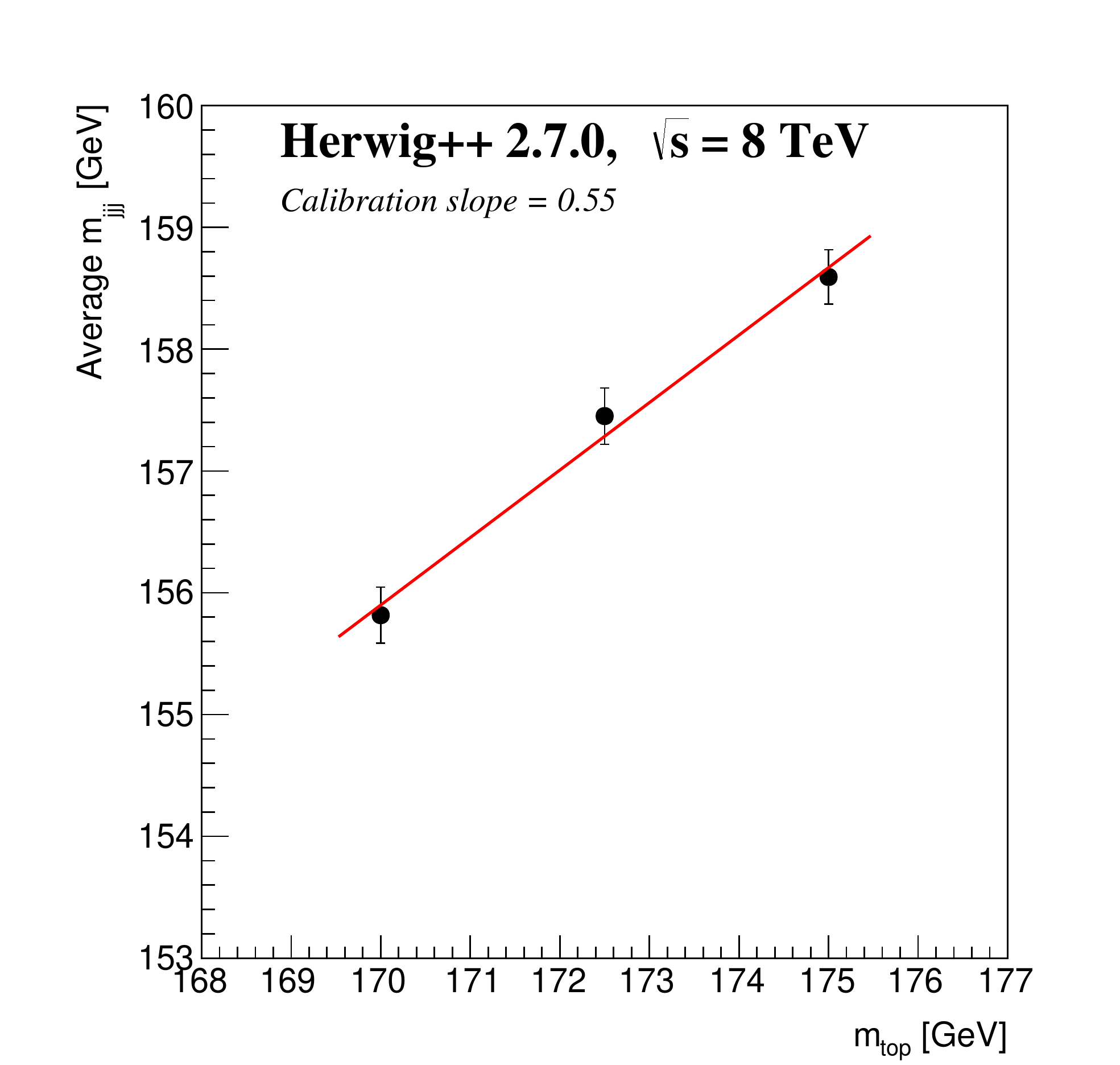}
\end{center}
\caption{Left: Unit normalized distributions of the $m_{jjj}$ variable for \ttbar\ with $m_t=172.5$\,\GeV, and for \tone\ pair production with a two-body \topLSP\ decay with $m_{\tone}=175$\,\GeV\ (and $m_t=172.5$\,\GeV), and a three-body decay \threeBody\ for $m_{\tone}=170$\,\GeV.  Right:  Calibration curve that relates the measured value $\langle m_{jjj}\rangle$ to the (MC) top quark mass, $m_t$.}
\label{fig:mjjj}
\end{figure}

The events are reconstructed using the {\sc Rivet 1.8.2} framework~\cite{rivet} and jets are clustered using {\sc Fastjet 3.0.6}~\cite{fastjet} with the anti-$k_t$ algorithm~\cite{antikt} and radius parameter $R=0.4$. Stable particles (excluding electrons and muons) with $p_T>500$\,\MeV\ and $|\eta|<5$ are clustered into jets.  Jets are $b$-tagged\footnote{We do not emulate an efficiency loss $\epsilon$ or mistag rate $m$.  Such effects do not have a big impact and are similar between signal and background.  So long as the two true b-jets are leading and subleading, the probability to choose a tagged jet which is not a true $b$-jet is $\sim 4(1-\epsilon) m\sim 1\%$. } by identifying $b$-hadrons from the Monte Carlo truth record within a $\Delta R=\sqrt{\Delta\phi^2+\Delta\eta^2}$ cone of $0.4$ of the jet axis.  
Events are selected which have a single electron or muon (lepton) in the final state in order to identify \ttbar\ decays where one of the $W$ bosons from the $t\rightarrow bW$ decays into leptons and the other decays hadronically.  We require at least four jets with $p_T>25$\,\GeV\ and at least two must be $b$-tagged.  Leptons are required to have $p_T>25$\,\GeV\ and be at least $\Delta R>0.4$ from any jet.  The missing transverse momentum is the negative of the vector sum of all stable particles within $|\eta|<5$.  Three jets $j_1,j_2,b_1$, exactly one with a $b$-tag ($b_1$), are associated with the hadronically decaying top quark by minimizing the following $\chi^2$-like estimator:  

	$$\chi^2 = \frac{(m_{j_{1}j_{2}b_{1}}-m_{b_2l\nu})^2}{(20 \GeV)^2}+\frac{(m_{j_1 j_2}-m_W)^2}{(10 \GeV)^2},$$

\noindent where $j_i$ are from the set of all non $b$-tagged jets with $p_T>25$\,\GeV, $b_1$ and $b_2$ are the highest $p_T$ $b$-tagged jets (not necessarily in order), $m_W\sim 80$\,\GeV, and the neutrino four-vector is determined from the missing transverse momentum in the $x$ and $y$ coordinates and by requiring $m_{l\nu}=m_W$ for the $z$ component.\footnote{The solution to $m_{l\nu}=m_W$ is quadratic in the neutrino $p_z$ and the value corresponding to the smaller $\chi^2$ is used.  In some cases, there is no solution to the quadratic equation in which case the neutrino $p_z$ is set to zero.  The neutrino is assumed to be massless.}  A variable sensitive to the top quark mass is then given by $m_{jjj}\equiv m_{j_1j_2b_1}$.  Figure~\ref{fig:mjjj} shows the distribution of $m_{jjj}$ for SM \ttbar\ production with $m_t=172.5$\,\GeV\ along with the same distribution for \tone\ pair production with a two-body \topLSP\ decay with $m_{\tone}=175$\,\GeV\ (and $m_t=172.5$\,\GeV), and a three-body decay \threeBody\ for $m_{\tone}=170$\,\GeV.  In all SUSY scenarios considered, the lightest neutralino is assumed to be massless.  The SUSY distributions are significantly different than the one for SM \ttbar. For the three-body decay this is because of the lack of a resonant top quark.  Even for the two-body stop decay, which contains a resonant top quark, the distribution is shifted to slightly lower values due to the finite widths of both the stop and the top (the top quark Breit Wigner is skewed low).  

Due to the differences in kinematic distributions, the probability of passing the selection will also vary by process.  In the cases with a resonant top quark the acceptance for direct stop pair production is very similar to \ttbar, but the three-body model has a softer $p_T$ spectrum and so has a lower probability of passing the kinematic selection ($\sim 60\%$ lower).

One way of measuring the top quark mass is to measure the average value of $m_{jjj}$ in some window and then relate this average to the true top quark mass via simulation.  We use a window of $100$--$200$\,\GeV\ and the calibration curve which relates the measured value of $\langle m_{jjj}\rangle$ to the top quark mass $m_t$ is shown in the right plot of Fig.~\ref{fig:mjjj}. The measured top quark mass in the presence of a light stop is estimated using this technique, where the summed SM \ttbar\ and SUSY stop distributions in the $m_{jjj}$ observable is used considering the respective cross-sections and event selection acceptances. 
When varying the true (MC) top quark mass then this is done consistently in the SM and SUSY samples.
Note that the event selection acceptance is assumed to be independent of the beam energy and initial state (this is approximately true at the LHC).
Another way of measuring the top quark mass, which we have performed as a cross check, is to use a fit with a line-shape function which approximately describes the $m_{jjj}$ distribution in data. The fit is based on the {\sc RooFit} package~\cite{roofit} and employs a convolution of a Breit-Wigner and a Gaussian probability-density-function, with the same fit window of $100$--$200$\,\GeV, and using the fitted mean of the line-shape function as the top quark mass estimator. The resulting calibration curve has a slope of $\sim 0.7$.

\section{Results}\label{sec:results}

\begin{figure}
\begin{center}
\includegraphics[width=0.5\textwidth]{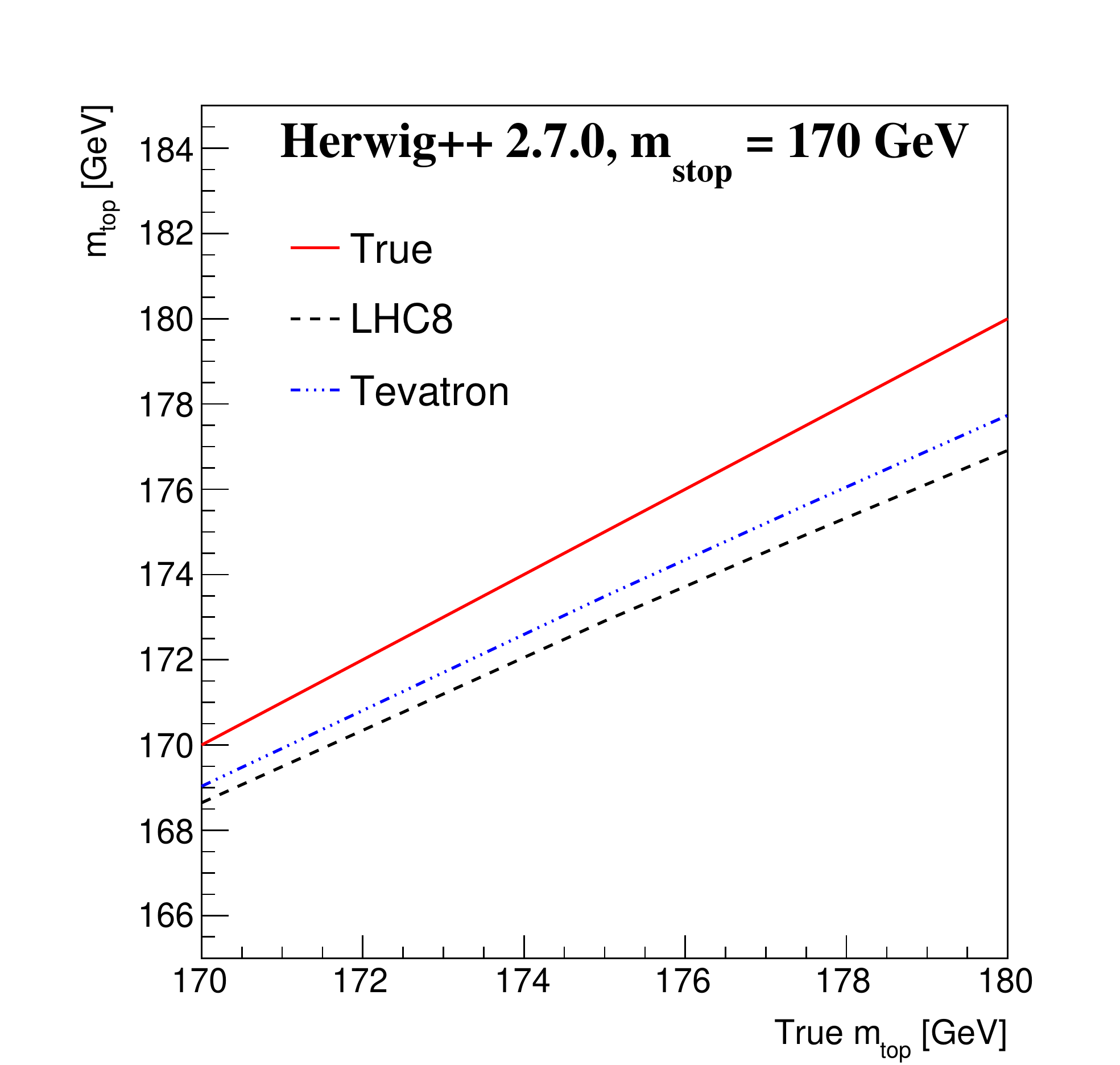}\includegraphics[width=0.5\textwidth]{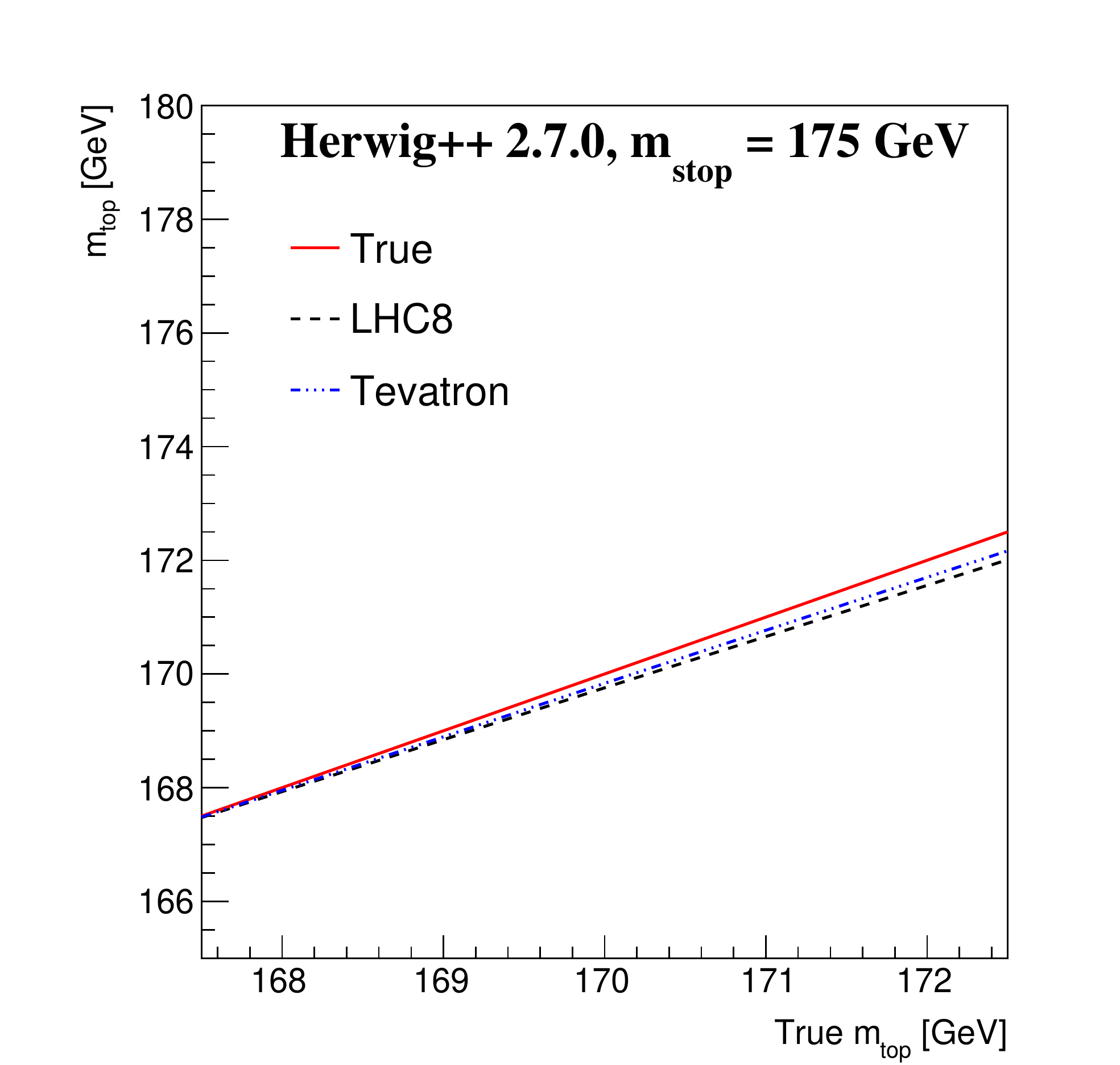}
\end{center}
\caption{The measured top quark mass as a function of the true top quark mass. The bias in the measurement arises from the presence of a light \tone\ with $m_{\tone}=170$\,\GeV\ and decaying via the three-body process (left) or with $m_{\tone}=175$\,\GeV\ and decaying via the two-body process.  
}
\label{fig:mass_shift}
\end{figure}

\begin{figure}
\begin{center}
\includegraphics[width=0.8\textwidth]{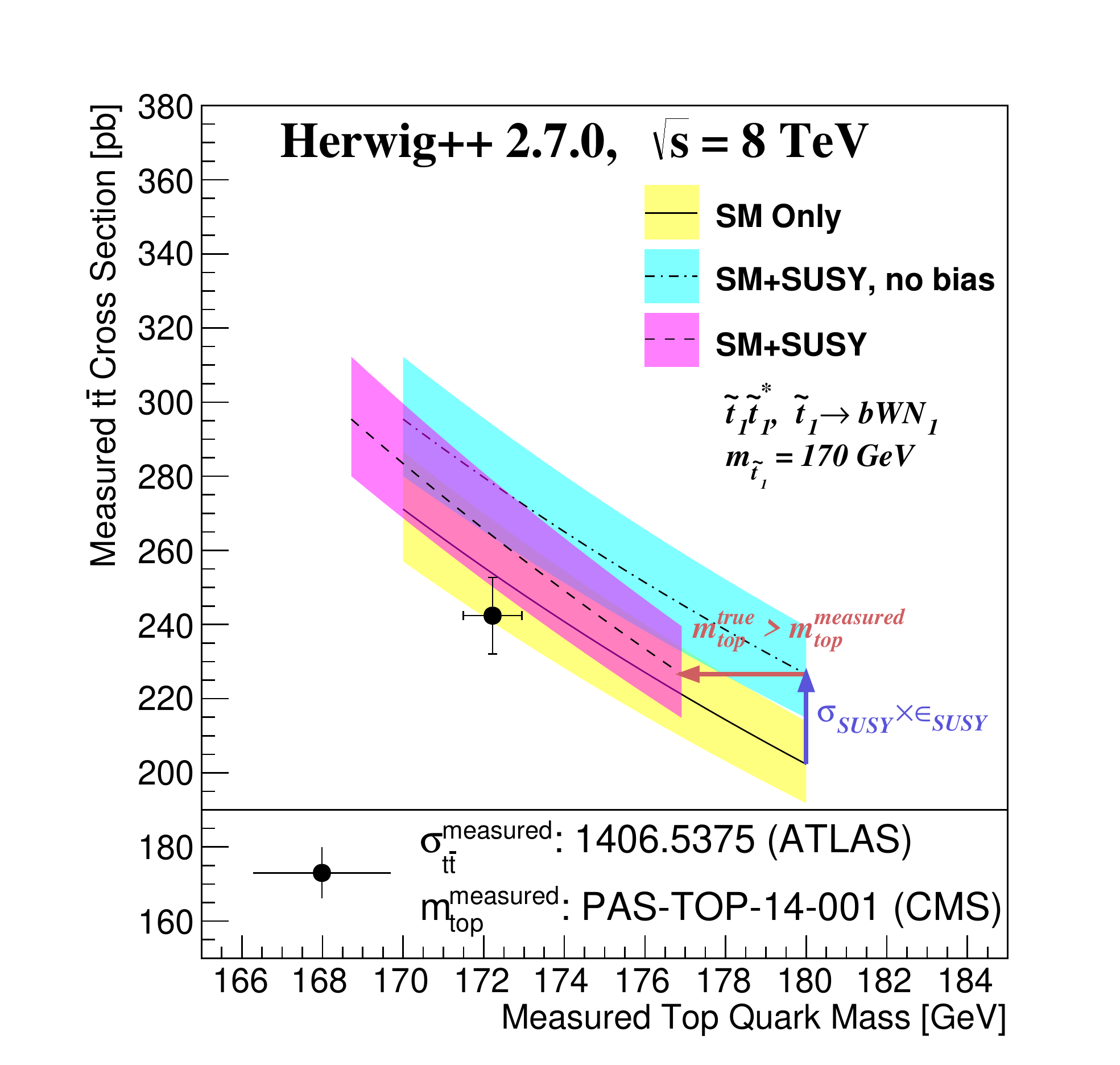}
\end{center}
\caption{
Summary of the effects leading to the sneaky stop scenario: the shifts in the measured \ttbar\ cross-section and measured top quark mass.  The solid line corresponds to an unbiased measurement of the \ttbar\ cross-section as a function of the top quark mass.  The dot-dashed line is what would be measured in the presence of a \threeBody\ with $m_{\tone}=170$ GeV for an unbiased top quark mass measurement.  However, under the SM+SUSY hypothesis the top quark mass measurement would be {\it biased} which translates into what would actually be observed shown in the dashed line.  For all three lines, the band reflects the $\sim 5-6\%$ theory uncertainty on the cross-section.  For comparison, the measured top quark mass and \ttbar\ cross-section are shown from recent CMS~\cite{cmsmass} and ATLAS~\cite{atlasxs} results.
}
\label{fig:summary}
\end{figure}

In general, the presence of a light stop that decays via the three- or two-body process reduces the measured top quark mass.  Figure~\ref{fig:mass_shift} shows the bias in the measured top quark mass as a function of the true top quark mass when including a light stop that decays either via the three- (left plot) or two-body process (right plot).   The shift due to two-body decays is much less than the impact of three-body stop decays.  Tables~\ref{tab:mass_shift}--\ref{tab:mass_shift2} list a selection of the numbers shown in Fig.~\ref{fig:mass_shift} and the corresponding impact on the measured cross-section.  Since the top quark mass would be measured too low, the predicted cross-section (based on the measured mass) would be too high, which can hide an excess of events due to stop pair production.  
For example, a stop with $m_\tone \sim 170$\,\GeV\ that decays via the three-body decay together with a true top quark mass of about $175$\,\GeV\ would cause
a bias in the top quark mass that makes it compatible with the measurements at the LHC8. As a consequence, the predicted \ttbar\ cross-section would be over-estimated by about $16$\,pb which in turn would make it much harder to find the stop with a cross-section of about $43$\,pb (which is further reduced to about $60$\% since the acceptance is lower than for \ttbar). 
The cross-section over-estimation increases with the true top quark mass, while the compatibility of the measured top quark mass with the LHC8 decreases when going beyond about $175$\,\GeV. 
Figure~\ref{fig:summary} summarizes how the change in the measured mass could hide a light stop decaying via the three-body process.  

There appears to be some tension between the top quark mass measured at the LHC8 and at the Tevatron; the difference of the two most-precise measurements (c.f. section~\ref{sec:method}) amounts to about $3$\,\GeV. The effect of a light stop biases the LHC8 more than the Tevatron, which would reduce the tension by about $0.6$\,\GeV\ in the above example. 
Turning this argument around to derive constraints on the presence of a light stop from the compatibility of the top quark mass obtained at different centre-of-mass energies (and/or $pp$ vs $p\bar{p}$) is currently precluded
on precision grounds.\footnote{We estimate the maximum top quark mass difference at the LHC8 and LHC14 in the presence of a light stop to be $0.3$\,\GeV\ or so.}

We have performed several additional checks to see how the results depend on the various method choices.  First, we have considered the dependance of the three-body bias on the stop and neutralino masses.  For fixed neutralino mass and lower stop mass, the distribution of $m_{jjj}$ shifts to lower values and 
the stop cross section increases, hence the bias increases (regulated by a small drop in acceptance due to softer $p_T$ spectra).  For fixed $\Delta_m=m_\tone- m_{N_1}$, the distribution of $m_{jjj}$ is roughly unchanged, but the bias can increase or decrease depending on the stop mass.  In particular, one could solve the following equations to find stop, neutralino, and top masses that are consistent with the measured values shown in Fig.~\ref{fig:summary}.  

\begin{align}
\label{eq:solve}
m_t^\text{measured} &= \frac{\langle m_{jjj}\rangle_{\stoppair}  \times \sigma_{\stoppair}(m_\tone)\times \epsilon+ \langle m_{jjj}\rangle_{\ttbar}(m_t) \times \sigma_{\ttbar}(m_t) }{ c_1(\sigma_{\stoppair}(m_\tone) \times \epsilon + \sigma_{\ttbar}(m_t))}-\frac{c_0}{c_1}\notag\\
 &\approx \frac{\langle m_{jjj}\rangle_{\stoppair} (\Delta_m)  \times \sigma_{\stoppair}(m_\tone)\times \epsilon(\Delta_m) + \langle m_{jjj}\rangle_{\ttbar} (m_t)  \times \sigma_{\ttbar}(m_t) }{ c_1(\sigma_{\stoppair}(m_\tone) \times \epsilon(\Delta_m) + \sigma_{\ttbar}(m_t))}-\frac{c_0}{c_1}\notag\\
\sigma_\text{$t\bar{t}$}^\text{measured} &= \sigma_{\stoppair}(m_\tone )\times \epsilon + \sigma_{t\bar{t}}(m_t) \notag\\
 &\approx \sigma_{\stoppair}(m_\tone )\times \epsilon(\Delta_m) + \sigma_{t\bar{t}}(m_t), 
\end{align}

\noindent where $\epsilon$ is the ratio of the SUSY acceptance to the $t\bar{t}$ acceptance and $c_0,c_1$ are the slope and intercept from the calibration curve in Fig.~\ref{fig:mjjj}, respectively.  One can approximate $\langle m_{jjj}\rangle_{\stoppair} \approx \langle m_{jjj}\rangle_{\stoppair} (\Delta_m)$ and $\epsilon \approx \epsilon(\Delta_m)$.
The nominal stop mass of $170$\,\GeV\ is already close to the optimal `hiding' point for the sneak stop, but the agreement with the measurement can be further improved by slightly increasing the stop mass for the assumed stop - neutralino mass difference.  

Two other additional checks are related to the theoretical an experimental modeling.  We have verified that the same qualitative shift in the $m_{jjj}$ distribution is observed when using a matrix element calculated with Madgraph~\cite{madgraph} at leading order interfaced with Herwig++ for the parton shower and hadronization.  We further checked that changing the experimental top mass measurement procedure does not qualitatively change the results.  Using the more sophisticated fit with a line-shape function described in section~\ref{sec:method} instead of $\langle m_{jjj}\rangle$, we still observe a significant bias for the three-body decays, though it is reduced by about $\mathcal{O}(10\%)$ depending on parameters.  

The actual bias in the top quark mass needs to be determined using proper top quark mass analyses and detector simulation. If the findings of this article are confirmed then  
analyses relying on the predicted \ttbar\ cross-section that set
exclusion limits on a light stop decaying via the three-body process
need to take this effect into account.
The exclusion limits in the recent ATLAS results in refs.~\cite{atlasxs,atlas-spin} are robust against this effect since only the stop two-body decay mode is considered for which we find no significant bias.
The spin correlation measurement should retain its sensitivity to a stop also for the three-body decay mode. However, the preliminary ATLAS stop exclusion limit based on the spin correlation analysis~\cite{atlas-spin} obtains about $30$\% of its sensitivity from the \ttbar\ cross-section constraint. 

\begin{table}
\begin{center}
{\small
\begin{tabular}{| c |c|c |c |c| c| c| c| c| c|c|}
\hline
$m_t^\text{true}$ & \multicolumn{2}{c|}{$m_t^\text{measured}$} & \multicolumn{2}{c|}{True $\sigma_{\ttbar}(m_t^\text{true})$}  & \multicolumn{2}{c|}{True $\sigma_{\ttbar}(m_t^\text{measured})$}&\multicolumn{2}{c|}{True $\sigma_{\stoppair}$}& \multicolumn{2}{c|}{Measured $\sigma_{\ttbar}$} \\
\hline
	& LHC8 & Tevatron & LHC8 & Tevatron & LHC8 & Tevatron & LHC8 & Tevatron &LHC8 & Tevatron  \\
  \hline
  \hline
  170 & 168.6 & 169.0 & 271.1  & 8.0 & 279.0 & 8.1& 42.6 & 0.87  & 295.4&8.5\\
  172.5 & 170.8 & 171.3  & 251.7 & 7.3& 264.4&7.6& 42.6 & 0.87&276.0 &7.8\\
  175 & 172.9 & 173.5  & 233.8 &6.8 & 249.7&7.2& 42.6 & 0.87&258.1 &7.3\\
  \hline
\end{tabular}
\caption{
Bias in the measured top quark mass and \ttbar\ cross-section due to the presence of a light stop ($m_\tone = 170$\,\GeV) that decays via the three-body process.
All masses are in \GeV\ and all cross-sections are in pb.  The measured top quark mass is biased low from the true mass which results in the true cross-section at the measured top mass,  true $\sigma_{t\bar{t}}(m_t^\text{measured})$ to be higher than the true cross-section at the true mass, true $\sigma_{t\bar{t}}(m_t^\text{true})$. The former quantity is what would be predicted under the SM-only hypothesis in the presence of the 170 GeV stop.  The measured $\sigma_{t\bar{t}}$ is the sum of true $\sigma_{t\bar{t}}(m_t^\text{true})$ and true $\sigma_{\stoppair}$, corrected for the lower acceptance for the three-body decay.}
\label{tab:mass_shift}
}
\end{center}
\end{table}

\begin{table}
\begin{center}
{\small
\begin{tabular}{| c |c|c |c |c| c| c| c| c| c|c|}
\hline
$m_t^\text{true}$ & \multicolumn{2}{c|}{$m_t^\text{measured}$} & \multicolumn{2}{c|}{True $\sigma_{\ttbar}(m_t^\text{true})$}  & \multicolumn{2}{c|}{True $\sigma_{t\bar{t}}(m_t^\text{measured})$}&\multicolumn{2}{c|}{True $\sigma_{\stoppair}$}& \multicolumn{2}{c|}{Measured $\sigma_{\ttbar}$} \\
\hline
	& LHC8 & Tevatron & LHC8 & Tevatron & LHC8 & Tevatron & LHC8 & Tevatron & LHC8 & Tevatron  \\
  \hline
  \hline
  170 & 169.8 & 169.8 & 271.1  & 8.0 & 273.7 & 8.0& 36.8 & 0.70 &304.8 &8.6\\
  172.5 & 172.0 & 172.2  & 251.7 & 7.3& 255.4&7.4& 36.8 & 0.70 &285.4 &8.0\\
  \hline
\end{tabular}
\caption{
Bias in the measured top quark mass and \ttbar\ cross-section due to the presence of a light stop ($m_\tone = 175$\,\GeV) that decays via the two-body process.
All masses are in \GeV\ and all cross-sections are in pb.  For more details, see the caption for Table~\ref{tab:mass_shift}.}
\label{tab:mass_shift2}
}
\end{center}
\end{table}

\section{Conclusions}\label{sec:conclusions}

We have argued that:

\begin{enumerate}
\item   The presence of a light \tone with  $m_{\tone} \sim m_t$ can bias the measured value of the SM top quark mass.   The size of the bias depends on the stop decay pattern and the stop mass, but in general the biased measurement is lower than the true mass.  For the three-body stop decay and $m_{\tone} \sim 170$\,\GeV, the shift in mass is significant compared to the current experimental precision.
\item   This negative shift in the measured mass combined with the increase in the predicted \ttbar\ cross-section (at the biased top quark mass) makes the relationship between measured cross-section and measured top quark mass similar to the SM-only relationship.  Thus, a {\it sneaky} light stop can evade detection from precision measurements. 
\end{enumerate}

The results presented here are obtained using truth-level studies and simplifying assumptions about the top quark mass methodology. 
If confirmed, however, this could mean that SUSY is well within the energy reach of the LHC.

\section*{Acknowledgements}

We would like to thank Michael Peskin for many useful discussions and feedback on earlier versions of the manuscript, as well as Jamie Boyd and Andreas Hoecker for their useful feedback that helped to improve the clarity of this article.  BN is supported by the NSF Graduate Research Fellowship under Grant No. DGE-4747 and also supported by the Stanford Graduate Fellowship.

\appendix

\end{document}